\documentclass[twocolumn,pre,aps,showpacs]{revtex4}
\usepackage{epsfig}
\begin{document}

\title{Scaling state of dry two-dimensional froths:\\
universal angle-deviations and structure}
\author{Andrew D. Rutenberg}
\email{andrew.rutenberg@dal.ca}
\homepage{http://www.physics.dal.ca/~adr}
\author{Micah B. McCurdy}
\affiliation{Department of Physics and Atmospheric Science, 
Dalhousie University, Halifax, Nova Scotia, Canada, B3H 3J5} 

\begin{abstract}
We characterize the late-time scaling state of 
dry, coarsening, two-dimensional froths using a detailed, 
force-based vertex model.  We find that the slow evolution of bubbles leads to systematic 
deviations from $120^\circ$ angles at three-fold vertices in the froth, with an amplitude 
proportional to the vertex speed, $v \sim t^{-1/2}$, but with a side-number dependence that is
independent of time.  We also find that a significant number of $T1$ side-switching processes 
occur for macroscopic bubbles in the scaling state, though most bubble annihilations involve
four-sided bubbles at microscopic scales. 
\end{abstract}



\date{\today}
\maketitle

\section{Introduction} 
\label{sec:intro}

Froth systems have a distinctive, close-packed cellular structure of bubbles that are bounded
by discrete smoothly curving walls (films) that intersect in point-like Plateau
borders (vertices).  Random soap froths continually coarsen due to gas 
diffusion between individual bubbles \cite{F105,F42,F19}.  
These dynamic froths exhibit a characteristic length-scale that grows as a power-law 
in time, $L \sim t^{1/2}$, as well as a time-independent ``scaling'' structure.
The scaling state, or regime, of froths is observed experimentally in both two-dimensional (2d) 
\cite{F46} and three-dimensional froths \cite{F40}.  Our understanding of the scaling state of $2d$ soap 
froths is largely built on the foundation of experimental \cite{F22,F100} and 
computational \cite{F43,F99,F104,F74,F14,F41} studies close of $2d$ froths, 
in the dry-froth limit with Plateau borders of vanishing size.

In the dry-froth limit, and under the approximation 
that bubble-wall curvatures and intersection-angles exactly satisfy local force- balance, 
individual bubble areas evolve with von Neumann's law,
\begin{equation}
	d A_n/d t = \pi \sigma D (n-6) /3,
	\label{EQ:vN}
\end{equation}
where $A_n$ is the area of a bubble with $n$ sides, 
$\sigma$ is the surface tension of the films, and $D$ is the coefficient of diffusion of 
gas between adjacent bubbles \cite{F19}.  This 
slow, continuous evolution is supplemented with dynamical rules for fast, topological 
rearrangements:  both ``$T2_n$'' processes of bubble annihilation for $n$-sided bubbles, 
and ``$T1$'' side-swapping processes between adjacent bubbles 
(see Fig.~\ref{T1T2fig} below). 

Experimental studies of undrained two-dimensional froths \cite{F46}
exhibit growing Plateau borders and systematic deviations of three-fold 
vertex angles from the force-balance angle of $2\pi/3$ \cite{F46}.
Conversely, experiments done in the dry-froth limit have not shown measurable
angle-deviations \cite{F22}.  The angle deviations reported for undrained
froths have been previously explained by the combination of film curvature and 
finite Plateau borders \cite{F32,F93,F109}.  However, small angle deviations 
are {\em required}, even for dry froths, in order to allow 
vertices to move and accommodate
continuous bubble growth.  In this paper, we use a natural force-based vertex model 
(see Sec.~\ref{SEC:forcebased}) that allows us to measure angle-deviations
for {\em dry} froths.  We find angle-deviations that are quantitatively consistent with 
experimentally observed angle deviations of undrained froths \cite{F46}.

Experimentally \cite{F57}, four- and five-sided bubbles have been observed to 
annihilate directly through $T2_4$ and $T2_5$ processes, without any intervening $T1$
processes, consistent with theory \cite{F25}.  However, previous numerical studies 
\cite{F43,F99,F104,F74,F14,F41,topological} have not clearly distinguished
between four- and five-sided bubbles that directly annihilate at microscopic scales 
and those that shed sides through $T1$ processes at macroscopic scales
before annihilating.  There is a related controversy over
whether any $T1$ events occur for macroscopic bubbles in the scaling regime
\cite{F41,F57}, or whether all late-time $T1$ events are associated with microscopic 
side-shedding during bubble annihilation \cite{F53,F107,F63}. 
In the scaling regime of our model froth, we confirm the experimental observations of 
direct annihilation of four- and five-sided bubbles at microscopic scales. We also 
observe a significant rate of macroscopic $T1$ events in the scaling state. 

\begin{figure}[htb]
\begin{center}
\includegraphics[width=250pt]{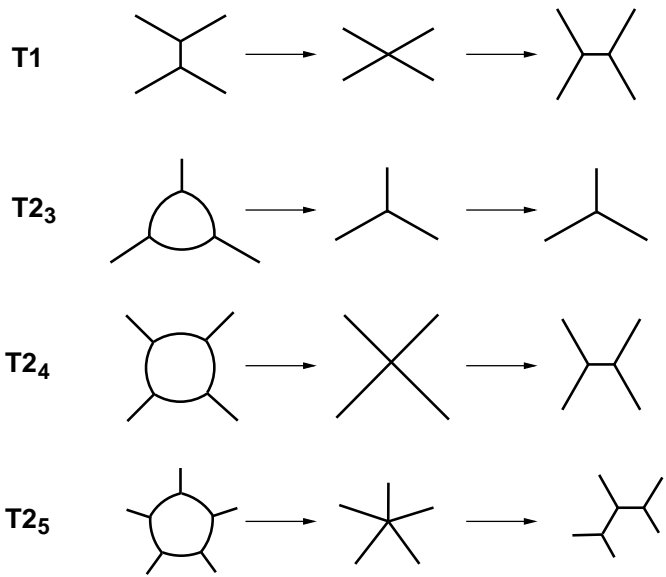}
\caption{Illustrations of the fast topological processes involved in froth evolution.
From top to bottom are a side-switching $T1$ event,
and three-, four-, and five-sided bubble deletions ($T2_3$, $T2_4$, and $T2_5$ respectively). 
From left to right are examples of the bubble walls and vertex positions before, during,
and after the corresponding topological process.  The 
details of resolution of $T2_4$ and $T2_5$ events will, in general, be 
determined by the forces acting along the 
bubble walls, as described in Sec.~\protect\ref{SEC:forcebased}. }
\label{T1T2fig}
\end{center}
\end{figure}

A standard measure of the topological structure in $2d$ froths is the second moment of
the side-distribution function, $\mu_2 \equiv \langle n^2 \rangle - \langle n \rangle^2$,
where $n$ is the number of walls of an individual bubble. 
Detailed froth models have recovered $\mu_2$ values
ranging from $1.2 - 1.6$ \cite{F74,F41,F14,F99} with reports of 
sensitive dependence of $\mu_2$ on the details of $T2_4$ and $T2_5$ processes \cite{F14,F74}. 
The frequently quoted experimental value of $\mu_2=1.4 \pm 0.1$ comes from an undrained 
froth experiment \cite{F46}.  Measurements of $\mu_2$ are bedeviled  by long-lasting initial transients. 
In this study, we find $\mu_2 = 1.24 \pm 0.01$ and show that the transient after the initial $\mu_2$ peak
can last a decade in time.  Our result is consistent with the late-time experimental results
of undrained froths \cite{F46}, and with numerical work on a locally equilibrated dry-froth \cite{F41}.

Computational model froths have either been fast but simple ``topological'' models 
(see, e.g., \cite{topological}), 
or slower but more detailed ``vertex'' models 
with vertex positions and film curvatures, 
\cite{F43,F99,F104,F74,F14,F41}.
With a natural force-resolved vertex model (section II), we 
characterize the scaling state of two-dimensional
dry froths. Starting with up to $10^4$ bubbles, we reach the scaling 
state after long transients.  In the scaling regime, we determine
the second-moment of the side-number distribution $\mu_2$ and 
the area growth-law amplitude $\langle A \rangle/(t D \sigma)$ 
(section III.A). 
We also characterize the 
angle deviations at bubble vertices, and find universal angle deviations 
associated with slow bubble coarsening (section III.B).  
Finally we find a significant
number of macroscopic $T1$ events occurring in the scaling regime, and find
that most bubbles annihilate as four sided bubbles through $T2_4$ events 
(section III.C).  

\section{Force-based Vertex model}
\label{SEC:forcebased}

We implement a fully force-based vertex model with natural deterministic 
dynamics for both the slow evolution of bubble areas and for fast 
topological rearrangements.  The motion of vertices is 
driven by the vector sum of the surface tension of the three films that meet 
at each vertex, leading to vertex velocities of 
\begin{equation}
	 \vec{v}  = \gamma \sigma \sum_i \hat \imath,
	\label{EQ:v}
\end{equation}
where the $\hat \imath$ are unit vectors tangent to the films meeting at a vertex 
and $\gamma$ is the vertex mobility.   We allow for film curvatures, $\kappa$, to 
dynamically relax towards their steady-state by 
\begin{equation}
	d \kappa/ dt = \Gamma (\Delta P - \kappa \sigma),
	\label{EQ:curve}
\end{equation}
where $\Gamma$ is the curvature mobility \cite{highcurve} 
and $\Delta P$ is the pressure difference between the 
two bubbles separated by the film.  Froth coarsening is driven by 
gas diffusion between bubbles with a rate proportional to the diffusion constant, $D$, the full length 
of the intervening film, and the pressure difference across the film: 
\begin{equation}
	d {\cal N}/dt = D \oint \Delta P dl, 
\end{equation}
where we impose an ideal gas relationship ${\cal N} = P A$, where $A$ is the bubble area and 
$P$ is the bubble pressure.   
The continuous evolution of vertices, curvatures, and bubble area is implemented with an 
Euler time-discretization with timestep $\Delta t$, supplemented with the dynamical 
rules for topological processes.  

The essence of both $T1$ and $T2$ processes is the temporary 
coalescence of two or more three-fold 
vertices into an $n$-fold vertex, followed by the dissolution of that $n$-fold vertex.  This 
all happens at microscopic length and timescales, determined in experimental systems by the 
Plateau border size.  Our approach implements this phenomenology, though we choose 
computationally convenient microscopic length-scales and rely on the 
universality of dry-froth structure to those scales.  We enter into $T1$ topological 
processes when adjacent vertices approach within a microscopic distance, $r_c$, of 
each other, and into $T2$ processes when a bubble area decreases 
below $10 r_c^2$. A finite $r_c$ allows us to use a fixed timestep, $\Delta t$ \cite{F41}. 
For a $T1$ process, we temporarily replace two participating three-fold vertices with one 
four-fold vertex at the midpoint, while, for $T2_n$ processes, we replace $n$ three-fold 
vertices with one $n$-fold vertex at the centroid of the deleting bubble.  
To resolve any $n$-fold vertex, we first 
determine the two adjacent films whose combined force upon the vertex is greatest,
in the center of force frame.   A short film of length $r_c$ is placed between 
these two films and the other films of the vertex, oriented with the combined force, 
giving an $n-1$-fold vertex and a 
three-fold vertex; this process is repeated until only three-fold vertices remain.  
This approach reproduces the natural instability of four- and five-fold vertices
\cite{F25}, and follows the lowest energy (highest force) channel for resolving $n$-fold
vertices.  The entire topological process is implemented between timesteps. 
Our results are not sensitive to $r_c$, as long as it is much
smaller than the average bubble scale, or to the precise 
details of the placement or resolution of the $n$-fold vertex at scales of $r_c$.  

Except where otherwise noted, we use $\sigma = 1$, $\gamma=1$, 
$\Gamma = 0.5$, and $D = 0.2$. 
We use a timestep $\Delta t = 0.01$ and $r_c = 0.01$.
Our systems are initialized with up to $N_0=10^4$ bubbles with a random but periodic
Voronoi construction. To control for finite-size effects, we have investigated different
$N_0$ while keeping a fixed initial average bubble area and pressure, $A_0=10^3$ 
and $P_0=1$, respectively.  We average at least $50$ independent samples for $N_0=10^4$, 
and displayed error bars are statistical \cite{fitrange}.

It is useful to compare our model to two recent vertex models,  one
by Herdtle and Aref \cite{F41} and another by Chae and Tabor \cite{F14}.
Herdtle and Aref build their model around an exact implementation of 
von Neumann's law, Eqn.~\ref{EQ:vN}, in the late-time limit
and neglect any effects of angle-deviations at vertices or of film-curvatures. 
This exact implementation requires computational power of $O(N^{1.3})$ for each timestep \cite{F41}, 
where $N$ is the number of bubbles, as opposed to $O(N)$ for our algorithm.  In terms of our 
dynamics, they implement the $\gamma,\  \Gamma \rightarrow \infty$ limit in Eqn.~(\ref{EQ:v}) 
and (\ref{EQ:curve}), respectively.  They only allow
$T2_3$ bubble deletions, and resolve shrinking four- and five-sided bubbles very finely in time
to catch all of the deterministic $T1$ side-shedding that occurs at microscopic scales 
\cite{F41}.  Chae and Tabor also impose curvature equilibrium, i.e. $\Gamma \rightarrow \infty$, but
have taken a length-dependent vertex mobility $\gamma \propto 1/ \sum l_i$ where $l_i$ are the lengths
of the films adjoining the vertex.   At late times, as typical side-lengths grow as 
$l \sim t^{1/2}$, they will approach the $\gamma=0^+$ limit --- which we expect to be singular. 
They allow all $T2_n$ processes, but impose various 
side-shedding rules for resolving $T2_4$ and $T2_5$ processes --- and find that 
$\mu_2$ depends on which {\em ad hoc} scheme is used \cite{F14}.   Like these two vertex
models, our 
model is also effectively in the high curvature mobility limit compared to the {\em natural} 
curvature dynamics \cite{highcurve} and
indeed we find that curvatures quickly approached the equilibrated limit $\kappa= \Delta P/\sigma$ as 
$t \rightarrow \infty$ (data not shown).   We believe that our force-based resolution of 
$T1$ and $T2_n$ events is equivalent to that of Herdtle and Aref \cite{F41} 
since the same film asymmetries that lead to side-shedding at microscopic scales in their 
model will lead to vertex shedding from $n$-fold vertices in our model. 

\section{Results}
\label{sec:results}

\subsection{Scaling State}
\label{subsection:universal}

Two-dimensional soap froths exhibit dynamic scaling at late times.
In the scaling state, the structural properties
of the froth are time-independent once appropriately adjusted for the shrinking number
of bubbles $N(t)$ and growing average area per bubble $A(t)$.  Topological structure such as 
the distribution function of side-number of individual bubbles becomes 
time-independent without scaling. 
Correlations in cellular systems, such as $2d$ froths, appear to
be universal due to the strong separation in length- and time-scales between the 
continuous evolution of area and the fast
$T1$ and $T2_n$ processes. Dry $2d$ froth experiments at various temperatures 
recover universal correlations \cite{F100}, as do undrained
experiments that vary the gas phase and hence diffusivity, $D$ \cite{F46}. 
Our results are consistent with the existence of a universal scaling state
at late times, though we have not systematically explored parameter space. 

\begin{figure}[htb]
\begin{center}
\includegraphics[width=250pt]{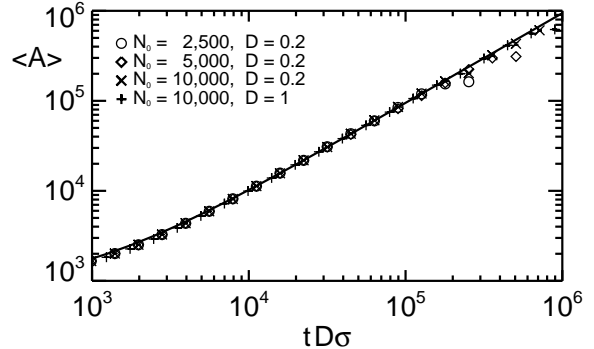}
\caption{The average bubble area of a coarsening froth, $\langle A \rangle$, 
vs. the natural time variable $\tilde{t} \equiv t D \sigma$. We see that linear growth is obeyed at 
late times, with a best-fit amplitude $\langle A \rangle = 0.94^{\pm 0.01} (\tilde{t}-\tilde{t}_0)$ indicated
by a solid line \protect\cite{fitrange}. 
This amplitude is expected to universally apply to coarsening froths in their scaling state, with 
initial conditions determining the effective time origin $\tilde{t}_0$. The same symbol legend applies
to other figures.}
\label{PrefactorFig}
\end{center}
\end{figure}

From von Neumann's law, we can directly see that $\langle A \rangle \sim t$ \cite{F100},
and this is generally observed \cite{F107,F52,F19}.  Since the 
froth evolution is driven by von Neumann's law (Eqn.~\ref{EQ:vN}), the natural time-variable is 
\begin{equation}
	\tilde{t} \equiv t D \sigma ,
\end{equation} 
and includes both the diffusivity and the surface 
tension, but not the curvature mobility, $\Gamma$\cite{highcurve}. 
As shown in Fig.~\ref{PrefactorFig}, we 
recover the expected coarsening law for $\tilde{t} \gtrsim 10^4$, after initial transients.
We see finite-size effects enter for $\tilde{t} \gtrsim 2 \times 10^5$ for $N_0=10^4$ ---roughly when 
the number of remaining bubbles drops below $100$.  We 
fit \cite{fitrange} the amplitude $\langle A \rangle /\tilde{t} = 0.94 \pm 0.01$ in Fig.~\ref{PrefactorFig} --- 
which we expect to be universal.    

\begin{figure}[hbt]
\begin{center}
\includegraphics[width=250pt]{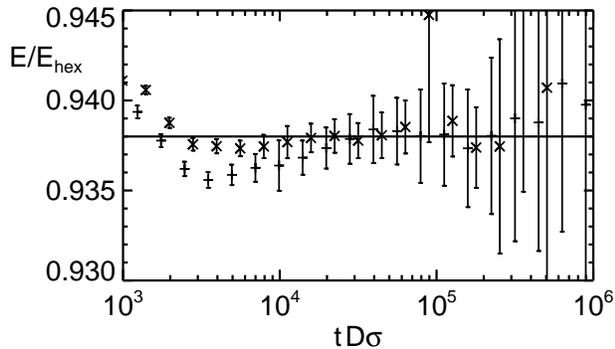}
\caption{Film energy of a coarsening froth, $E$, equal to the surface tension $\sigma$ times the total
film length, normalized by the energy of a uniform hexagonal froth 
$E_{\mathrm{hex}} \equiv N \sigma\sqrt{2\sqrt 3 A_0}$,
with the same number of bubbles, $N$, and average bubble area, 
$A_0$, plotted vs. $\tilde{t} = t D \sigma$. The asymptotic value in the scaling state is 
$E/E_{\mathrm{hex}} = 0.938 \pm 0.001 $ \protect\cite{fitrange} indicated by the horizontal line.} 
\label{LineLengthFig}
\end{center}
\end{figure}

We next measure the surface energy of the froth, $E$,
normalized by $E_{\mathrm{hex}} = N \sigma\sqrt{2\sqrt 3 A_0}$,
the energy of a regular hexagonal froth with the same average bubble
size, $A_0$ \cite{F41,F14}. Our result \cite{fitrange}, $E/E_{\mathrm{hex}} = 0.938 \pm 0.001$ 
is shown in Fig.~\ref{LineLengthFig} and is consistent with the best previous measurement
($0.945 \pm 0.010$ \cite{F41}), though with higher precision.  

\begin{figure}[htb]
\begin{center}
\includegraphics[width=250pt]{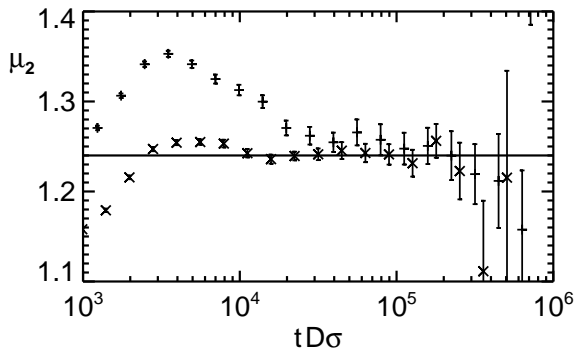}
\caption{The second moment of the side-number distribution, $\mu_2$ vs $\tilde{t}$. 
We extract \protect\cite{fitrange} 
an asymptotic value of $\mu_2 =1.24 \pm 0.01$, indicated by a horizontal line.
Significant transients are seen for the larger $D$ data (``$+$''), including a distinctive
peak. }
\label{Mu2fig}
\end{center}
\end{figure}

Simulations of dry froths \cite{F99,F91,F74,F14,F41} 
have given values of the second-moment of the distribution of the number of sides in a 
bubble, $\mu_2 \equiv \langle n^2 \rangle -\langle n \rangle^2$, 
that range between $1.2$ and $1.6$.  As shown in Fig.~\ref{Mu2fig}, 
we observe \cite{fitrange} a constant, asymptotic value of $\mu_2 = 1.24 \pm 0.01$.
While broadly consistent with previous values, we are 
closest to Herdtle and Aref \cite{F41}, who found $\mu_2 \approx 1.2$. 

While we expect that $A/\tilde{t}$, $E/E_{\mathrm{hex}}$, and $\mu_2$ are universal
measures of the scaling state, 
both the initial froth configuration and the particular dynamical parameters of 
the model will affect initial transients.  By varying $D$, we can simulate the effects of using various
gas-phase components --- much as experimental studies have used either helium or air in their froths 
\cite{F46}.  For larger $D$, we qualitatively reproduce a strong pre-asymptotic $\mu_2$  peak in
Fig.~\ref{Mu2fig} that was seen
experimentally \cite{F46}.  The transient regime after that $\mu_2$ peak 
lasts approximately {\em one decade} in time.  In the best experimental determination of
$\mu_2$, the entire experiment lasted only one decade after the $\mu_2$ peak \cite{F46}, and the 
latest time data also recovered $\mu_2 \approx 1.2$.

\subsection{Universal Angle Deviations}
\label{SEC:angledev}

In a force-balanced steady-state, the angles between the three adjacent films
that form a vertex will all be $\theta_0 \equiv 2 \pi/3$. Any deviations from 
$\theta_0$ will result in a net force and motion of the vertex due to Eqn.~(\ref{EQ:v}). Since 
most bubbles in the froth are continuously growing due to Eqn.~(\ref{EQ:vN}), their
vertices must be continuously moving --- necessitating non-zero angle deviations.
We expect internal angles typically greater than $\theta_0$ for growing bubbles ($n>6$)
and internal angles less than $\theta_0$ for shrinking bubbles ($n<6$). 

\begin{figure}[htb]
\begin{center}
\includegraphics[width=250pt]{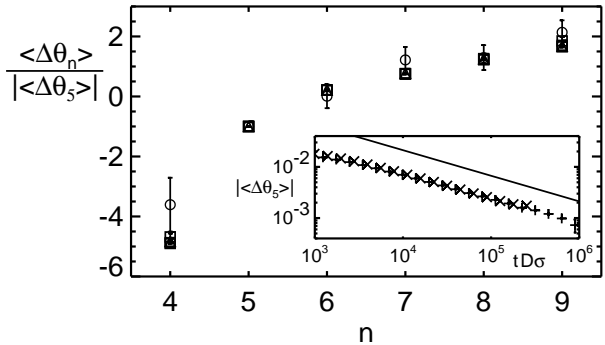}
\caption{Average angle deviations of $n$-sided bubbles scaled by 
the $n=5$ sided deviations, 
$\langle \Delta \theta_n \rangle/ |\langle  \Delta \theta_5 \rangle|$, vs. $n$.  
The data is for $\tilde{t}=27969$ (triangles and squares) 
and $\tilde{t}=158217$ (``$\times$'' and
``$+$''), for $D=0.2$ and $D=1$, respectively.  Also shown (circles) 
is the scaled data from experimental undrained froths \protect\cite{F46}, 
which illustrates the universality of the scaled angle deviations. Inset:
time dependence of angle deviations of $n=5$ sided bubble vertices, 
$|\langle \Delta \theta_5 \rangle|$, scaled by $\gamma D^{-1}$ (see text) 
vs $\tilde{t} \equiv t D \sigma$.  
The solid line shows the expected $\tilde{t}^{-1/2}$ behavior.}
\label{AngleDevfig}
\end{center}
\end{figure}

In the scaling regime, the average bubble area grows as 
$\left<A\right> \sim \tilde{t}$, with a characteristic length
$L \equiv \sqrt{\left<A\right>} \sim \sqrt{\tilde{t}}$ and vertex speed 
$v \sim dL/dt \sim  \sigma D \tilde{t}^{-1/2}$. 
Small angle-deviations will, by Eqn.~(\ref{EQ:v}), be
proportional to vertex speeds divided by $\gamma \sigma$, so we expect
$\left<\left|\Delta \theta\right|\right> \sim D \gamma^{-1} \tilde{t}^{-1/2}$, where 
$\Delta \theta \equiv \theta-\theta_0$.  As shown in the inset of Fig.~\ref{AngleDevfig}, we
observe this time-dependence in the scaling regime.  We also collapse the data for different
$D$ by scaling by $\gamma D^{-1}$.

Different size Plateau borders will lead to different vertex mobilities, 
$\gamma$, and will in turn require different magnitude angle deviations for the same 
vertex speeds.  If we scale out the magnitude, 
we should be left with universal angle deviations that reflect the universal froth structure. 
Indeed, in our data we find that angle deviations as a function of the number of bubble sides, 
scaled by the value for $n=5$, is time-independent in the scaling regime for $3 < n<10$ 
(see Fig.~\ref{AngleDevfig}).  Underlining this universality, the experimental data from 
undrained froths \cite{F46}, indicated by circles in the figure, is consistent with our results 
within error bars.  Angle deviations have been previously explained as arising from hidden
angles due the finite size of 
Plateau borders in combination with finite film curvatures \cite{F32} 
(see also \cite{F93,F109}).  This explanation is unsupported by our data. 
Our angle deviations are measured at the vertex center and
do not include any explicit effects due to Plateau border size. 

Topological events also contribute to locally substantial 
angle deviations that are, however, sub-dominant in the scaling regime. 
We can qualitatively understand these deviations 
under the simplifying assumption that all angles equal $\theta_0$ immediately
before a topological event.  As illustrated in Fig.~\ref{T1T2fig}, 
immediately after a $T1$ event, two of the six internal angles of the two separating 
vertices are $\pi/3$ rather than $2 \pi/3$. Similarly, immediately after a $T2_4$ 
event two of the six internal angles about two vertices are $\pi/4$, 
and immediately after a $T2_5$ event two of the 
nine internal angles around three vertices are $\pi/5$.  These large angle
deviations will last until a quasi-static configuration is reached by 
moving vertices on order the bubble size, $L \sim t^{1/2}$.
As shown in the next subsection, the rate of topological events is proportional 
to the rate of bubble annihilation, 
$R_{T2} \sim - dN/dt  \sim t^{-2}$, so the average angle deviation due to
topological events is expected to be only 
$\left<\left|\Delta \theta\right|\right>_{T} \sim t^{-3/2}$, which is sub-dominant to 
the effects of continuous bubble evolution.  

\subsection{Topological Processes}
\label{SEC:topo}

It is known that three-, four-, and five-sided bubbles can annihilate directly 
through $T2_3$, $T2_4$, and $T2_5$ 
processes respectively \cite{F25}, however it has not been clear what the relative rates 
of these processes are.  To complicate the matter, 
while a three-sided bubble can only annihilate through a $T2_3$ process, a four- or five-sided 
bubble can shed sides through $T1$ processes at macroscopic scales 
and then annihilate through a $T2_3$ process.
There have been qualitative reports that there are few if any 
$T1$ processes that are not associated with side-shedding
on the way to annihilation \cite{F107,F57}, 
and, indeed, simulations that only allow $T2_3$ processes  are 
self-consistent and yield results comparable to other models \cite{F41}.  
However, $T2_4$ and $T2_5$ processes have been reported experimentally \cite{F57}.  

What has been lacking is a clear definition of a $T1$ process as distinct from a $T2$ process. 
A natural distinction is between macroscopic $T1$ processes and microscopic $T2$ processes. 
$T1$ processes should be counted distinctly if they are associated {\em only} with macroscopic bubbles. 
Any side-shedding that occurs at microscopic scales should simply be counted as the details of the associated
$T2$ process.  We have a natural microscopic scale that corresponds to the Plateau border size, $r_c$, 
while the average bubble size, $L \sim t^{1/2}$, provides an appropriate macroscopic scale. 
The geometric mean of these two lengths provides a natural division between microscopic and macroscopic, 
$\ell \equiv \sqrt{r_c L(t)}$.
At late times, $\ell$ will be far from both microscopic and macroscopic scales.  We count
$T1$ events distinctly if all bubbles involved are bigger than $\ell^2$, and 
count $T2_n$ events if the bubble has $n$ sides as it shrinks past 
the scale of $\ell^2$.  Scaling $\ell$ by a constant factor can significantly 
change the duration of initial transients, but it should not not change results 
in the scaling regime.  

\begin{figure}[htb]
\begin{center}
\includegraphics[width=250pt]{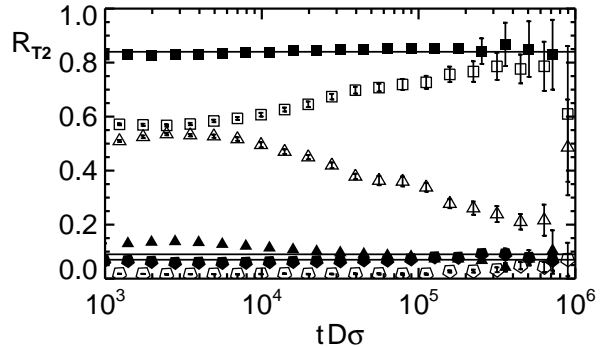}
\caption{The proportion of $T2_3$, $T2_4$, and $T2_5$ processes, for three, four, and five-sided
bubble annihilation, denoted by triangles, squares, and pentagons, respectively, 
vs. $\tilde{t}$. The filled symbols are for $D=0.2$ while the open symbols are for
$D=1$, both with $N_0=10^4$.  The number of sides are counted as
the bubble area approaches microscopic scales, at $\ell \equiv \sqrt{r_c L(t)}$. 
The proportions approach constant values of $9^{\pm 2}\%$, $84^{\pm 2}\%$, and $7^{\pm 2}\%$ 
respectively, for $D=0.2$ \protect\cite{fitrange}, indicated by horizontal lines. 
Significant transients are apparent even at late times for $D=1$, though the trend is towards
agreement with $D=0.2$. }
\label{345Fig}
\end{center}
\end{figure}

We have measured the proportion of bubbles which vanish with three, four, and five sides, through 
the $T2_3$, $T2_4$, and $T2_5$ processes illustrated in Fig.~\ref{T1T2fig}. The proportions of these
processes have not been reported in the literature, despite qualitative reports of $T2_4$ and $T2_5$
processes in undrained froths \cite{F57}.  As shown in Fig.~\ref{345Fig}, 
these proportions are constant in the scaling regime, and have the values 
of $9^{\pm 2}\% $, $ 84^{\pm 2}\%$, and $7^{\pm 2} \%$ for $T2_3$, $T2_4$, and $T2_5$ processes, respectively. 
What is surprising is how few bubbles annihilate with three-sides.  This is qualitatively
consistent with how few small three-sided bubbles are observed in froths in the scaling
state;  it also indicates how important $T2_4$ and $T2_5$ processes
are in froth dynamics.

\begin{figure}[htb]
\begin{center}
\includegraphics[width=250pt]{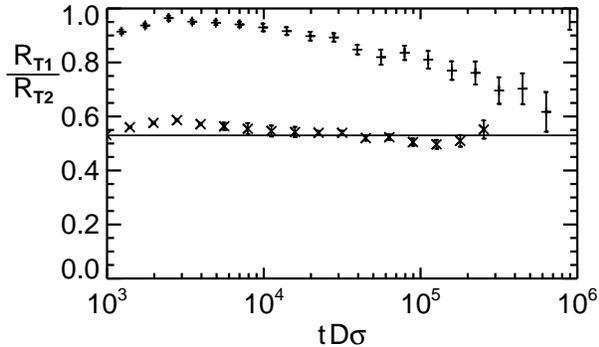}
\caption{The proportion $R_{T1}/R_{T2}$ of macroscopic $T1$ events to microscopic $T2$ events, where
only side-shedding involving bubbles larger than $\ell \equiv \sqrt{r_c L(t)}$ are counted as $T1$ events.
We find a ratio of $R_{T1}/R_{T2}=0.53 \pm 0.05$ in the scaling regime \protect\cite{fitrange},
indicated by the horizontal line. Very long transients are found for $D=1$ (``$+$''), consistent with the 
previous figure.}
\label{TopFreqFig}
\end{center}
\end{figure}

We have also measured the ratio of rates of macroscopic $T1$ to bubble annihilation $T2$ processes, 
$R_{T1}/R_{T2}$.  This ratio has been measured in simulations by Herdtle and Aref \cite{F41} to 
be $3/2$, but they included  microscopic side-shedding 
events so their result provides an upper bound to ours.  
As shown in Fig.~\ref{TopFreqFig}, we find a ratio of $R_{T1}/R_{T2} \approx 0.53 \pm 0.05$. 
This indicates that $T1$ events are significant even in late stages of froth coarsening.  

\section{Summary and Discussion}
\label{sec:summary}

We have accurately characterized the scaling regime of a two-dimensional
dry froth with natural force-resolved topological processes. We find 
values of $\mu_2 = 1.24 \pm 0.01$, $\langle A \rangle/\tilde{t} = 0.94 \pm 0.01$,
and $E/E_{\mathrm{hex}} = 0.938 \pm 0.001$, independent of the gas diffusivity $D$.  Our results 
for $\mu_2$ and $E/E_{\mathrm{hex}}$ are consistent with previously measured values \cite{F41}
though considerably more accurate, while our results for $\langle A \rangle/\tilde{t}$ have 
not been previously reported.   These results are consistent with the hypothesis of a universal
scaling state in two-dimensional froths, independent of the microscopic details of the model
as long as natural topological processes are used \cite{F14}.
To affect the structure, dynamical features should either change the asymptotic von Neumann's law
or the resolution of topological processes. So froth rupture \cite{F14} and shear \cite{shear}
should and do affect structure, but varying the experimental temperature, the plateau border size, 
or the initial conditions do not. We expect that our basic froth model yields universal structure in 
the scaling state if $D$, $\gamma$, $\Gamma$, and $\sigma$ are all non-zero, however a systematic 
exploration has not yet been done.  From the precision of our results, it now appears feasible to undertake 
such a study.  We have shown that deviations from $2 \pi/3$ vertex angles have a non-universal 
magnitude $\langle|\Delta \theta| \rangle \sim t^{-1/2}$ that is a natural 
consequence of slow bubble growth and finite vertex mobility, $\gamma$. 
When scaled by the angle deviations for $n=5$ bubbles, the scaled angle deviations are time-independent
and agree quantitatively with experiment \cite{F46} --- indicating their universality. This 
universality reflects the universality of the froth structure, as well as the robustness of the
topological processes to the small angle-deviations that remain in the scaling-state. 
While the angle deviations appear to reflect rather than affect the scaling structure
of the froth, this may not be the case in sheared froths \cite{shear}. Angle deviations must be
quite large to affect the resolution of $T1$ processes, but even small deviations can 
affect the resolution of $T2_4$ and $T2_5$ processes.  We therefore expect that angle 
deviations will significantly affect structure in rapidly sheared coarsening froths.  
Furthermore, while the angle deviations following a topological process are sub-dominant in 
coarsening froths, they may be relevant to sheared froths since in those systems topological 
processes are not well-separated in time at late times. 

We have also characterized the topological processes occurring in the scaling state. We make a
clear distinction between $T2$ events occurring at microscopic scales, on the order of $r_c$, and 
$T1$ events occurring at macroscopic scales, on the order of $\sqrt{\langle A \rangle}$.  We
find that most annihilating bubbles have four-sides ($84^{\pm 2}\%$)
while relatively few have three ($9^{\pm 2}\%$) or five sides ($7^{\pm 2}\% $).   
We also characterize the ratio of side-switching $T1$ events to bubble-annihilating $T2$ events
and find a significant number of $T1$ events for macroscopic bubbles in the scaling state, 
$R_{T1}/R_{T2} = 0.53 \pm 0.05$.  This is a significant result, since many topological froth models 
neglect macroscopic $T1$ processes.  This simplification appears
to be unjustified.  We would expect small but significant effects in the scaling structure due to these
$T1$ processes.  It will be interesting to block macroscopic $T1$ events within our dynamics, 
and compare the resulting froth structure with our results.

Our agreement with the results of Herdtle and Aref \cite{F41}, 
indicates that our natural force-based resolution of $T2$ processes is consistent with 
their deterministic side shedding within $T2$ processes at microscopic scales.  This also 
indicates that their large vertex mobility limit, $\gamma \rightarrow \infty$, recovers the 
universal scaling state of the coarsening froth. Our agreement with the undrained froth
experiments of Stavans and Glazier \cite{F46}, including the scaled angle deviations, 
indicates that the small corrections to von Neumann's law due to Plateau border broadening 
does not appear to change the structure of the scaling state significantly. 

Our force-based vertex model is simple, versatile, fast, and accurate.  It should be 
straightforward to study froths with $N_0$ up to $10^6$ in the near future, including the 
effects of shear.  This will allow a more detailed investigation of universal scaling-state 
structure in soap-froths. 

\acknowledgments

We thank the National Science and Engineering Research Council of 
Canada for support. ADR thanks Petro-Canada for a Young Innovator award for this research.
Computer resources were provided by the Canadian Foundation
for Innovation, by the Institute for Research in Materials at Dalhousie University, 
and also by the Advanced Computational Research Laboratory at the 
University of New Brunswick.  We thank Michael Greenwood and Peter Cordes for initial
coding and stimulating discussions. 



\begin{thebibliography}{99}
\bibitem{highcurve} The natural curvature dynamics consistent with the microscopics
	for an isolated bubble is $v_\perp= \lambda^{-1}(\Delta P - \kappa \sigma)$ \protect\cite{F19} 
	where $\lambda$ is an inverse mobility. This  
	leads to curvature dynamics 
	$\dot{\kappa} \propto \lambda^{-1} \kappa^2 (\Delta P - \sigma \kappa)$, which implies
	that at late times, as $\kappa \rightarrow 0$,  
	our dynamics, Eqn.~\protect\ref{EQ:curve}, is in the high-mobility or
	$\lambda \rightarrow 0^+$ limit. We have used this limit in Eqn.~\protect\ref{EQ:vN} 
	\protect\cite{F19}.
\bibitem{fitrange} Fit values are extracted from an error-weighted average of 
the $D=0.2$ data for size $N_0 = 10^4$ systems in the interval $\tilde{t} \in [2 \times 10^4,2 \times 10^5]$. 
Reported errors on fit values include estimated systematic effects.
\end{thebibliography}
\end{document}